\documentclass[]{aa}  

\usepackage{graphicx}
\usepackage{color}
\usepackage{txfonts}

\begin{document}

   \title{The ALMA-PILS survey: propyne (CH$_3$CCH) in IRAS 16293--2422}
   \titlerunning{Propyne (CH$_3$CCH) in IRAS 16293--2422}

   \author{H. Calcutt\inst{1}, E. R. Willis\inst{2}, J. K. J{\o}rgensen\inst{3}, P. Bjerkeli\inst{1}, N. F. W. Ligterink\inst{4}, A. Coutens\inst{5}, H. S. P. M\"uller\inst{6}, R. T. Garrod\inst{2}, S. F. Wampfler\inst{4}, M. N. Drozdovskaya\inst{4}}
\institute {
\inst{1}Department of Space, Earth and Environment, Chalmers University of Technology, 41296, Gothenburg, Sweden, \email{hannah.calcutt@chalmers.se}\\
\inst{2}Departments of Chemistry and Astronomy, University of Virginia, Charlottesville, VA 22904, USA\\
\inst{3}Niels Bohr Institute, University of Copenhagen\\ \phantom{'}{\O}ster Voldgade 5--7, DK-1350 Copenhagen K., Denmark\\
\inst{4}Center for Space and Habitability, University of Bern, Gesellschaftsstrasse 6, CH-3012 Bern, Switzerland\\
\inst{5}Laboratoire d'Astrophysique de Bordeaux, Univ. Bordeaux, CNRS, B18N, all\'{e}e Geoffroy Saint-Hilaire, 33615 Pessac, France\\   
\inst{6}I. Physikalisches Institut, Universit\"at zu K\"oln, Z\"ulpicher Str. 77, 50937 K\"oln, Germany\\
}

\authorrunning{Calcutt, H., Willis, E. R. et al.}
   \date{Received}
 
  \abstract
   {Propyne (CH$_3$CCH), also known as methyl acetylene, has been detected in a variety of environments, from Galactic star-forming regions to extragalactic sources. Such molecules are excellent tracers of the physical conditions in star-forming regions, allowing the temperature and density conditions surrounding a forming star to be determined.}
   {This study explores the emission of CH$_3$CCH in the low-mass protostellar binary, IRAS 16293--2422, examining the spatial scales traced by this molecule, as well as its formation and destruction pathways.}
   {Atacama Large Millimeter/submillimeter Array (ALMA) observations from the Protostellar Interferometric Line Survey (PILS) are used to determine the abundances and excitation temperatures of CH$_3$CCH towards both protostars. This data allows us to explore spatial scales from 70 to 2400 au. This data is also compared with the three-phase chemical kinetics model MAGICKAL, to explore the chemical reactions of this molecule.  }
   {CH$_3$CCH is detected towards both IRAS 16293A and IRAS 16293B, and is found to trace the hot corino component around each source in the PILS dataset. Eighteen transitions above 3$\sigma$ are detected, enabling robust excitation temperatures and column densities to be determined in each source. In IRAS 16293A, an excitation temperature of 90 K and a column density of 7.8$\times$10$^{15}$ cm$^{-2}$ best fits the spectra. In IRAS 16293B, an excitation temperature of 100 K and 6.8$\times$10$^{15}$ cm$^{-2}$ best fits the spectra. The chemical modelling finds that in order to reproduce the observed abundances, both gas-phase and grain-surface reactions are needed. The gas-phase reactions are particularly sensitive to the temperature at which CH$_4$ desorbs from the grains.}
   {CH$_3$CCH is a molecule whose brightness and abundance in many different regions can be utilised to provide a benchmark of molecular variation with the physical properties of star-forming regions. It is essential when making such comparisons, that the abundances are determined with a good understanding of the spatial scale of the emitting region, to ensure that accurate abundances are derived.}


   \maketitle

\section{Introduction}
Complex organic molecules (COMs) are excellent tracers of the physical conditions in star-forming regions owing to the large number of transitions that span a range of upper energy levels. When they are found in dense environments, they are easily thermalised and can be an excellent probe of the specific temperature conditions surrounding a forming star. One such molecule, propyne (CH$_3$CCH or CH$_3$C$_2$H), also known as methyl acetylene, is well-suited for such studies. It is a prolate symmetric top molecule which means that for a given J quantum number, there is a K-ladder which has a large range of energy levels, emitting over a narrow frequency range. It is also relatively abundant in many star-forming regions, and has therefore been detected in a variety of environments from Galactic star-forming regions (e.g. \citealt{Snyder1973, Bogelund2019}), to extragalactic sources \citep{Martin2006, Harada2018}. It is increasingly becoming an important molecule for large comparison studies of chemical diversity among star-forming regions (e.g. \citealt{Taniguchi2018}), especially with its ability to link the chemistry of extragalactic sources to local environments, and even planetary atmospheres (e.g. Titan; \citealt{Cordiner2015}).
 
Crucially, for CH$_3$CCH to function as a good reference species in a variety of sources, its column densities and excitation temperatures must be determined robustly. This can pose a significant problem for some sources, which have a very complex physical structure. Observations must be of sufficient angular resolution to ensure that the scales over which CH$_3$CCH is found are determined. Too high a resolution risks at least some of the emission being filtered out, whereas too low a resolution means beam dilution effects becoming a significant issue when determining column densities and excitation temperatures. 

One such observationally complex source is IRAS 16293--2422 (hereafter IRAS 16293). IRAS 16293 is a low-mass protostellar binary, located at a distance of 141\,pc \citep{Dzib2018}. A recent survey of IRAS 16293, the Protostellar Interferometric Survey (PILS; \citealt{Jorgensen2016}), has detected more than 100 molecules from $\sim$10,000 lines, highlighting the rich molecular chemistry that can occur in hot corino sources. Its two Class 0 protostars, A and B, are separated by a distance of 5\arcsec\ ($\sim$700\,au) and exhibit a rich chemistry in each of their respective hot corinos.  IRAS 16293A has a large velocity gradient across a disk-like structure with a near to edge-on morphology, whereas IRAS 16293B has a face-on morphology showing little velocity gradient across the source \citep{Pineda2012,Jacobsen2018}. IRAS 16293A has even been suggested to be a proto-binary system, with multi epoch continuum observations showing that it contains two continuum peaks \citep{Chandler2005,Pech2010,Hernandez2019}. An overview of the literature on this source can be found in \citet{Jorgensen2016}.

Propyne has been detected in IRAS 16293 previously by \citet{Cazaux2003} with the IRAM 30\,m telescope, and was attributed to the cold gas component, with an excitation temperature of 25\,K. Recent work by \citet{Andron2018}, with further IRAM 30\,m observations, constrained the emission of propyne to the outer part of envelope, at around 2000\,au and an excitation temperature of 25\,K, based on radiative transfer modelling of their observations. 

In this paper, the emission of propyne in both IRAS 16293A and IRAS 16293B is investigated, using higher angular resolution data than has previously been used. The results from this observational analysis are then compared to the results of chemical modelling of the IRAS 16293 system. The observations are presented in Section \ref{sec:obs}, and details of the results and analysis are presented in Section \ref{sec:res}. In Section \ref{sec:mod}, details of the chemical modelling are given and in Section \ref{sec:diss} the wider impact of the results from this study are discussed. Finally, the conclusions are presented in Section \ref{sec:cons}. %

\section{Observations}\label{sec:obs}
The observations used in this study are part of the Protostellar Interferometric Line Survey (PILS), an ALMA spectral line survey of IRAS 16293 \citep{Jorgensen2016}. The survey covers Band 7 between 329.147 and 362.896\,GHz, and selected frequency windows in Bands 3 and 6. For the purposes of this study the Band 7 and Band 3 data are used. The Band 7 data are a combination of 12\,m array and Atacama Compact Array (ACA) observations, which have been combined to have a restoring beam of 0\farcs5 at a spectral resolution of 0.2\,km\,s$^{-1}$. These data probe scales from 70\,au (0\farcs5) to 1820\,au (13\arcsec). The phase centre of the observations is located between the two components of the binary system at $\alpha_{J2000}$ = 16$^{\rm h}$32$^{\rm m}$22{\rm \fs}72, $\delta_{J2000}$=$-$24$^{\circ}$28\arcmin34\farcs3. They reach a sensitivity of about 7\,--\,10\,mJy\,beam$^{-1}$\,channel$^{-1}$, i.e. approximately $\sim$5\,mJy\,beam$^{-1}$\,km\,s$^{-1}$ across the entire frequency range. The Band 3 data cover four basebands, spanning the frequency ranges 89.5\,--\,89.7\,GHz, 92.8\,--\,92.9\,GHz, 102.5\,--\,102.7\,GHz, and 103.2\,--\,103.4\,GHz. They probe spatial scales of 1.4\arcsec\ (196\,au) to 17\arcsec\ (2380\,au) and reach a sensitivity of $\sim$5\,mJy\,beam$^{-1}$\,km\,s$^{-1}$. The spectral resolution of the data is 0.4\,km\,s$^{-1}$. Further details of the PILS datasets and the reduction and the continuum subtraction procedure can be found in \citet{Jorgensen2016}. \\

\section{Observational results and analysis}\label{sec:res}

The PILS data has been used to search for CH$_3$CCH and its isotopologues towards both IRAS 16293A and IRAS 16293B. Spectra were extracted from two positions, where line emission is bright, does not have strong absorption features and does not suffer from high continuum optical depth (\citealt{Coutens2016}, \citealt{Lykke2017}). The first position is at $\alpha_{J2000}$=16$^{\rm h}$32$^{\rm m}$22{\rm \fs}58, $\delta_{J2000}$=$-$24$^{\circ}$28\arcmin32\farcs8. This is offset from the continuum peak position of IRAS 16293B in the south west direction by one beam  (0\farcs5, 70\,au). The second position is at $\alpha_{J2000}$=16$^{\rm h}$32$^{\rm m}$22{\rm \fs}90, $\delta_{J2000}$=$-$24$^{\circ}$28\arcmin36\farcs2. This is offset from the peak continuum position of IRAS 16293A by 0\farcs6 (85\,au), in the north east direction. Both these positions have been used previously to study the chemistry in both hot corinos (e.g. \citealt{Ligterink2017}, \citealt{Calcutt2018b}, \citealt{Manigand2019}) to overcome issues of line blending and absorption contamination which affect the peak continuum position. \\

Rest frequencies and related information on CH$_3$CCH, CH$_2$DCCH and CH$_3$CCD were taken from the CDMS catalogue \citep{Endres2016}. The CH$_3$CCH data are based on \citet{Cazzoli2008}, \citet{Muller2000} and \citet{Muller2002}. The CH$_2$DCCH and CH$_3$CCD data are largely based on \citet{LeGuennec1993}, with most of the data coming from that source.

A total of 18 lines of CH$_3$CCH are detected above 3$\sigma$ towards both IRAS 16293A and IRAS 16293B. There are 5 lines that are not severely blended towards IRAS 16293A and 14 lines that are not severely blended towards IRAS 16293B. There are more blended lines in IRAS 16293A because the lines are wider in this source. $^{13}$C and deuterated isotopologues of CH$_3$CCH were searched for but were not detected in the data. Upper limit column densities for these isotopologues are given in Table \ref{tab:uplimits}. The upper limits are determined from $1.05{\times}3{\times}RMS{\times}\sqrt{{\Delta}V{\times}FWHM}$ to compute the 3$\sigma$ limit, where 1.05 is a factor to account for a 5\% flux calibration uncertainty. Line parameters are used for the brightest lines of each isotopologue that fall in the range of the observations. Line widths of 2\,km\,s$^{-1}$ and 1\,km\,s$^{-1}$ are used for IRAS 16293A and IRAS 16293B, respectively. The CH$_3$CCH excitation temperature determined in this work is used for the upper limit calculations. The upper limit column densities values are found to be consistent with the isotopic ratios that have been determined in both sources by previous works (e.g. \citealt{Calcutt2018b, Jorgensen2018}). The spectroscopic information for the CH$_3$CCH lines that are detected is given in Table \ref{tab:spec}. \\

\begin{table*}[h!]
\caption{Spectroscopic information for the lines of CH$_3$CCH detected in IRAS 16293.}\label{tab:spec}
\centering
\begin{tabular}{ccccccc}
\hline
\hline
Transition&Frequency &$E\rm{_{u}}$ &$A\rm{_{ij}}$&$T_{\rm peak}$ IRAS 16293A& $T_{\rm peak}$ IRAS 16293B\\
&(GHz)&(K)&(s$^{-1}$)&K&K\\
\hline 
6 3 -- 5 3 &	102.53035&	82&	2.67$\times$10$^{-6}$&\phantom{1}5.54&10.10\\
6 2 -- 5 2 &	102.54015&	46&	3.16$\times$10$^{-6}$&\phantom{1}4.93&\phantom{1}8.66\\
6 1 -- 5 1 &	102.54602&	24&	3.46$\times$10$^{-6}$&\phantom{1}6.76&11.50\\
6 0 -- 5 0 &	102.54798&	17&	3.56$\times$10$^{-6}$&\phantom{1}7.49&12.60\\
 20 6 -- 19 6 & 341.50702& 432& 1.26$\times$10$^{-4}$&\phantom{1}1.66&\phantom{1}4.48\\
 20 5 -- 19 5 & 341.57846& 353& 1.30$\times$10$^{-4}$&\phantom{1}2.06&\phantom{1}5.09\\
 20 4 -- 19 4 & 341.63695& 288& 1.34$\times$10$^{-4}$&\phantom{1}4.28&\phantom{1}9.66\\
 20 3 -- 19 3 & 341.68247& 237& 1.36$\times$10$^{-4}$&14.10&28.10\\
 20 2 -- 19 2 & 341.71499& 201& 1.38$\times$10$^{-4}$&10.90&21.60\\
 20 1 -- 19 1 & 341.73451& 179& 1.39$\times$10$^{-4}$&13.70&26.00\\
 20 0 -- 19 0 & 341.74102& 172& 1.39$\times$10$^{-4}$&14.70&27.60\\
 21 6 -- 20 6 & 358.57236& 449& 1.48$\times$10$^{-4}$&\phantom{1}1.54&\phantom{1}4.22\\
 21 5 -- 20 5 & 358.64734& 370& 1.52$\times$10$^{-4}$&\phantom{1}1.90&\phantom{1}4.77\\
 21 4 -- 20 4 & 358.70873& 305& 1.55$\times$10$^{-4}$&\phantom{1}3.93&\phantom{1}9.07\\
 21 3 -- 20 3 & 358.75650& 254& 1.58$\times$10$^{-4}$&13.00&26.60\\ 
 21 2 -- 20 2 & 358.79063& 218& 1.60$\times$10$^{-4}$&10.10&20.30\\
 21 1 -- 20 1 & 358.81112& 197& 1.61$\times$10$^{-4}$&12.60&24.50\\
 21 0 -- 20 0 & 358.81795& 189& 1.61$\times$10$^{-4}$&13.60&26.10\\
\hline
\end{tabular}

\end{table*}
\subsection{Column densities and excitation temperatures}
\begin{figure*} 
\begin{center} 
\includegraphics[width=18.6cm, angle=0, clip =true, trim = 7cm 0cm 5cm 0cm]{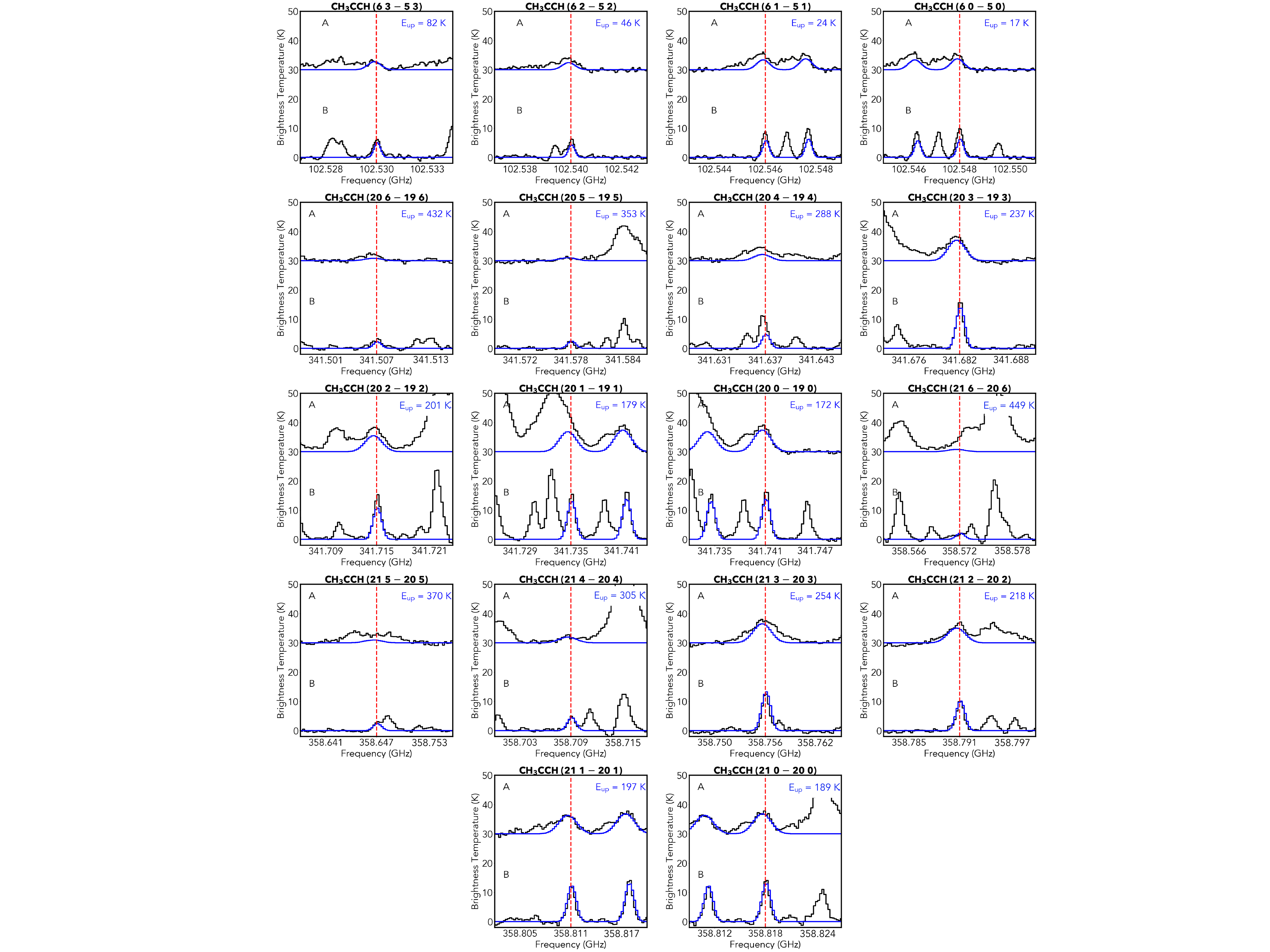}\\

 \end{center} 
  \vspace{-0.5cm}
\caption{The lines of CH$_3$CCH detected in IRAS 16293A and IRAS 16293B (black) overlaid with an LTE spectral model (blue). The upper energy level of each line is given in the top right corner of each plot in K. IRAS 16293A spectra are plotted with an offset of 30 K on the same axis as IRAS 16293B spectra. The red dashed line indicates the systemic velocity for each source.\label{fig:spectra}} 
 \end{figure*} 
In order to determine the best fit column densities and excitation temperatures for CH$_3$CCH towards both sources, a local thermodynamic equilibrium (LTE) model was fitted to the data. For the large densities seen in the environments close to the two protostars, LTE is a good approximation. The spectral modelling software CASSIS\footnote{CASSIS has been developed by IRAP-UPS/CNRS: \url{http://cassis.irap.omp.eu/}.} was used to calculate synthetic spectra and determine best fit spectral models, by running a grid of models covering a range of column densities, excitation temperatures, full width at half maximums (FWHM), and $V_{\rm peak}$ values. Column densities were varied between 5$\times$10$^{13}$\,--\,5$\times$10$^{18}$\,cm$^{-2}$, and excitation temperatures between 80\,--\,300 K. FWHM values were varied between 2.0\,--\,2.6\,km\,s$^{-1}$ for IRAS 16293A and 0.8\,--\,1.2\,km\,s$^{-1}$ for IRAS 16293B. $V_{\rm peak}$ values were varied between 0.8\,--\,1.2\,km\,s$^{-1}$ for IRAS 16293A and 2.5\,--\,2.7\,km\,s$^{-1}$ for IRAS 16293B. The range in FWHMs and peak velocities modelled was based on previous analysis towards each of the offset positions used in this work (e.g. \citealt{Calcutt2018b,Manigand2019}). This range covers several velocity channels either side of the previous best fit values to account for lines that might be broader or narrower than previously studied molecules, and molecules which may have a different peak velocity. The reduced $\chi^2$ minimum was then computed to determine the best fit model, assuming a source size of 1\farcs4 and 1\farcs1 for A and B respectively, for the higher upper energy lines in Band 7, and a source size of 1\farcs6 and  1\farcs7 for A and B respectively, for the lower upper energy lines in Band 3. These source sizes are based on fitting a 2D Gaussian to the emission, which is discussed in Section \ref{sec:spaex}. The significant spectral density seen in IRAS 16293 means that line blending is an issue within the spectra for both sources A and B. Lines are considered to be severely blended, if the maximum intensity of a line overlaps with the first minimum (i.e. where the intensity of the line falls to zero) of a nearby line \citep{Snyder2005}. Such lines are excluded from the best fit model calculations. Lines of CH$_3$CCH fall within the range of both the Band 3 and Band 7 datasets. Initially, a fit was performed to each dataset independently, and then a combined dataset was modelled. All of the best fit parameters were similar for each source independent of which dataset was modelled. 

The flux calibration errors on the Band 7 dataset were estimated to be better than 5\% by \citet{Jorgensen2016}. Both the flux calibration error and the root mean square (RMS) are used by CASSIS when determining the best fit model. A larger source of error, however, comes from the quality of the fit of the spectral model to the observational data. Line blending and optically thick lines can significantly impact the quality of the fit. To mitigate this, first an initial grid of models was run on all of the emission lines, and then a second grid of models was performed, where  only the optically thin lines ($\tau<$0.2) were used to determine the best fit model. The errors on the fit parameters were then estimated by varying the $N_{\rm tot}$ and $T_{\rm ex}$ to determine the impact this had on the fit of the observations.

The results from this LTE modelling are given in Table \ref{tab:coldens} and the spectra and best fit model of the brightest lines in IRAS 16293A and IRAS 16293B are shown in Figure \ref{fig:spectra}. Similar excitation temperatures and abundances are found between both sources. Towards IRAS 16293A the best fit FWHM and peak velocity is 2.0 km\,s$^{-1}$ and 1.0 km\,s$^{-1}$ respectively. These values are consistent with previous studies performed at the offset position in IRAS 16293A used in this work (e.g. \citealt{Calcutt2018b}). Towards IRAS 16293B the fitted FWHM of 1.0 km\,s$^{-1}$ is the same as has been determined for other molecules analysed as part of the PILS survey (e.g. \citealt{Lykke2017}). The peak velocity is 2.5 km\,s$^{-1}$, which is only 1 channel width (0.2 km\,s$^{-1}$) different from the source velocity. Such a small shift in peak velocity was also seen in some oxygen-bearing species in \citet{Jorgensen2018}. They also found that molecules detected towards IRAS 16293B fell into two categories depending on the excitation temperature, `hot' ($\sim$300 K) and `cold' ($\sim$125 K), which correlate with the binding energy of the molecules to the grain-surface, i.e. `cold' molecules sublimate before `hot' molecules. CH$_3$CCH falls in the category of a `cold' molecule, which is consistent with its binding energy of 4287 K. This value is based on the binding energy of acetylene (C$_2$H$_2$), measured by \citet{Collings2004}, combined with the binding energy of 2 H atoms. Other estimates of the CH$_3$CCH binding energy by \citet{Behmard2019} find it to have a value between 4400\,--\,4700K.

\begin{table*} 
\caption{Excitation temperature ($T_{\rm ex}$), column density ($N\rm{_{tot}}$), the abundance ratio, N(CH$_3$CCH)/N(H$_2$), $FWHM$, and peak velocity ($V_{\rm peak}$) in IRAS 16293B and IRAS 16293A.}\label{tab:coldens}
\centering
\footnotesize
\begin{tabular}{cccccc}
\hline
\hline
Source&$T\rm{_{ex}}$&$N\rm{_{tot}}$  &\underline{$N$(CH$_3$CCH)$^{\dagger}$}&$FWHM$&$V_{\rm peak}$\\
&(K)&(cm$^{-2}$)&$N$(H$_2$)&km\,s$^{-1}$&km\,s$^{-1}$\\
\hline
IRAS 16293A&90$\pm$30&7.8$\pm$1.0$\times$10$^{15}$&\phantom{x}1.2$\times$10$^{-9}$&2.0$\pm$0.2&1.0$\pm$0.2\\
IRAS 16293B&100$\pm$20&6.8$\pm$0.2$\times$10$^{15}$&$<$5.7$\times$10$^{-10}$&1.0$\pm$0.2&2.5$\pm$0.2\\
\hline

\end{tabular}
\tablefoot{$^{\dagger}$H$_2$ column densities are based on the values determined in \citet{Jorgensen2016} towards IRAS 16293B ($>$1.2$\times$10$^{25}\,$cm$^{-2}$) and values determined in \citet{Calcutt2018a} towards IRAS 16293A (6.3$\times$10$^{24}\,$cm$^{-2}$). }
\end{table*} 

\begin{figure*} 
\begin{center} 
\includegraphics[width=18.2cm, angle=0, clip =true, trim = 0cm 1cm 0cm 0cm]{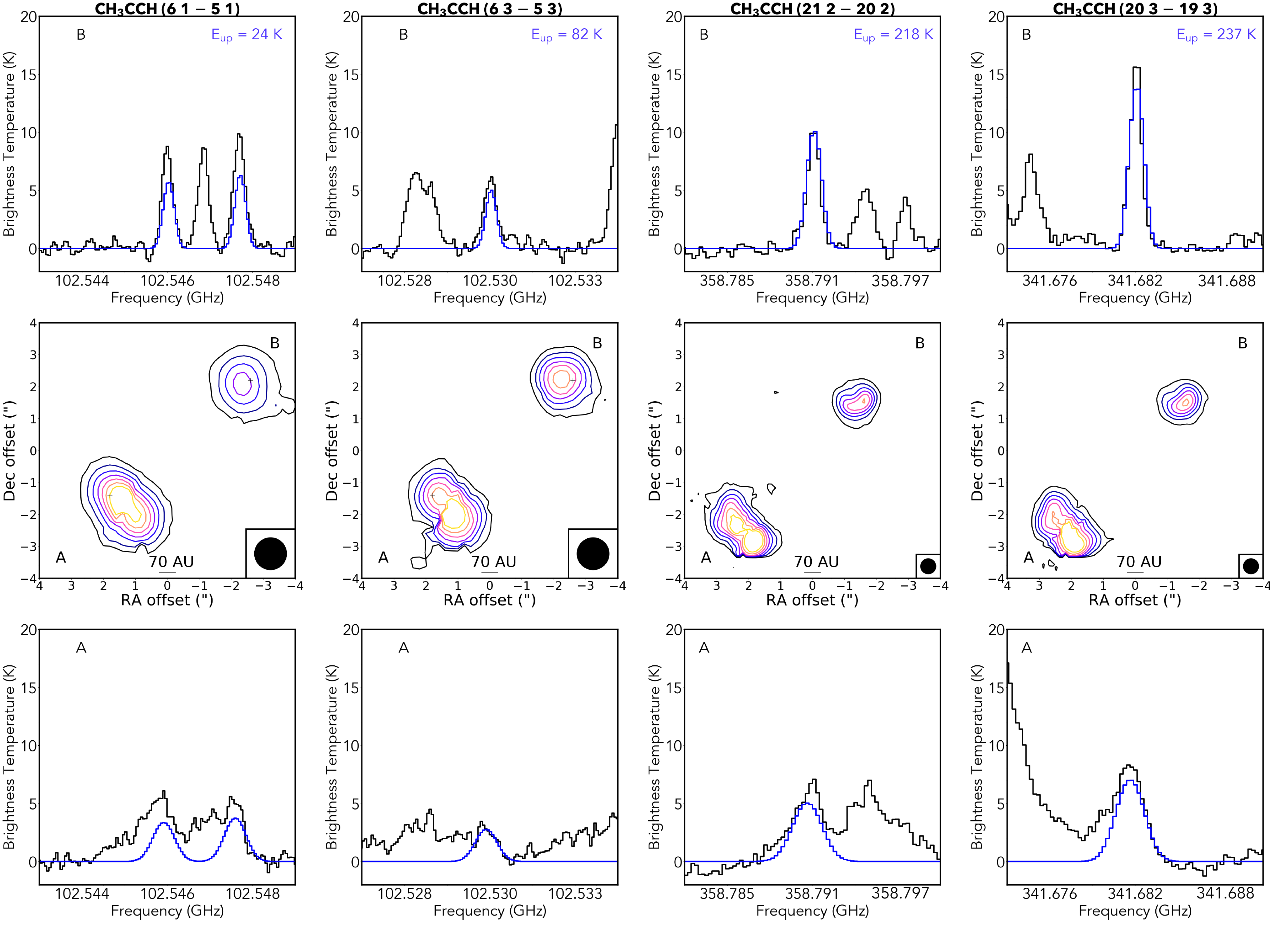}
\end{center} 
\caption{Emission maps and spectra of the four brightest CH$_3$CCH lines detected in both the Band 3 and Band 7 datasets. The panels are arranged in order of increasing upper energy level. The top panels show the spectra of the lines (black line) and LTE best fit model (blue line) towards IRAS 16293B and the bottom panel shows the same towards IRAS 16293A. The middle panel shows a velocity-integrated emission map (VINE) map of the emission towards each source. The velocity interval used is 3.2 km\,s$^{-1}$ and 1.2 km\,s$^{-1}$ for IRAS 16293 A and IRAS 16293B, respectively. The contours range from 10\%--90\% of the max flux in each plot in steps of 10\%. This corresponds to 8$\sigma$--63$\sigma$, 5$\sigma$--40$\sigma$, 18$\sigma$--144$\sigma$, and 20$\sigma$--167$\sigma$ for each of the middle panels. The RA and Dec coordinates are given relative to the phase centre of the observations which is the same in both the Band 3 and Band 7 datasets. The shift in position of both sources between the two datasets is to be expected due to the difference in time between the observations, and has been discussed previously (e.g. \citealt{Pech2010}). The upper energy level of each transition is given in the top right corner of the top panels. The beam size of the observations is given in the bottom right corner of the middle panels.\label{fig:ch3cch_combmap}} 
\end{figure*}

\begin{figure} 
\begin{center} 
\includegraphics[width=9.5cm, angle=0, clip =true, trim = 0cm 0cm 0cm 0cm]{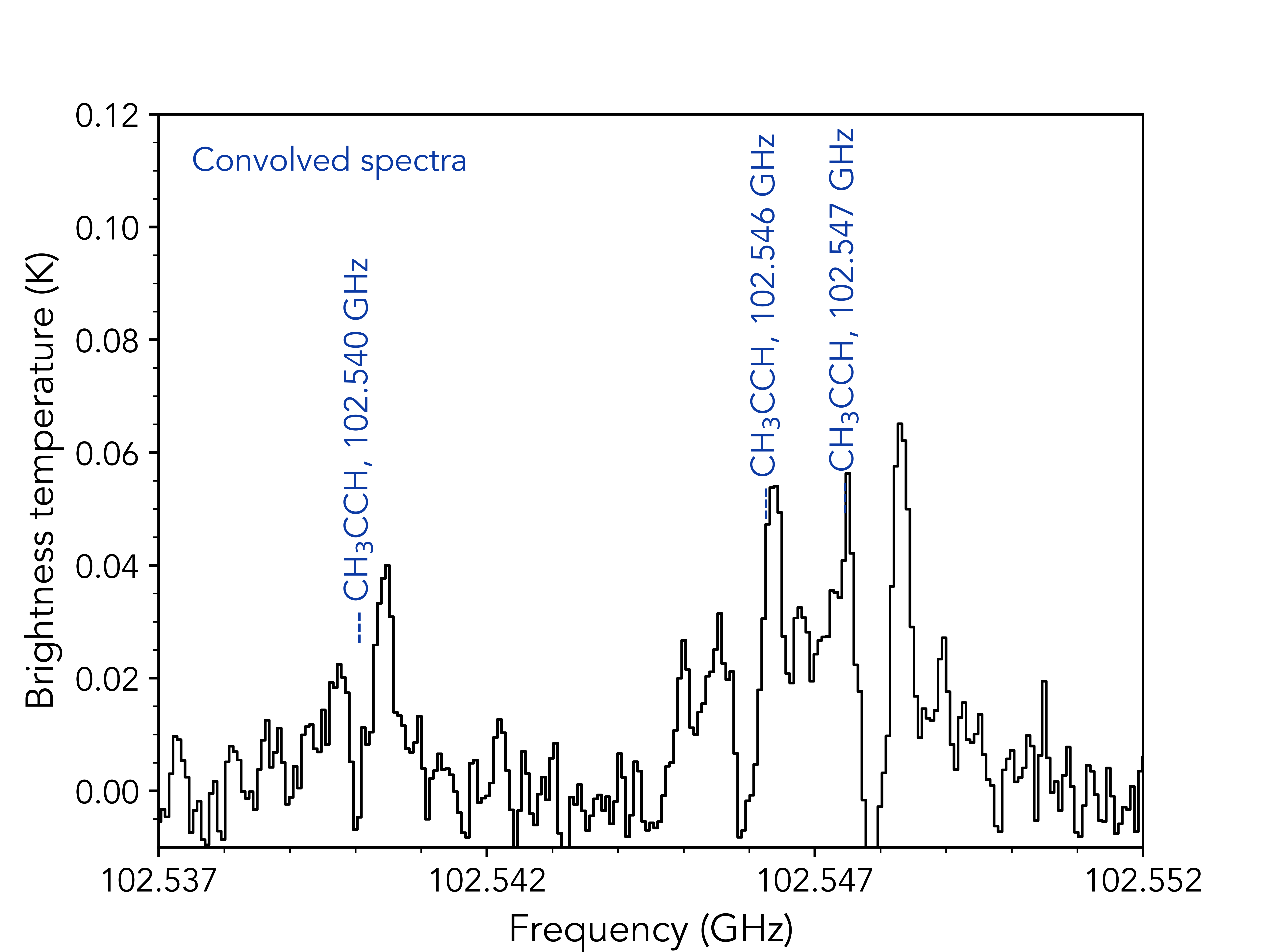}\\
 \end{center} 
  \vspace{-0.5cm}
 \caption{Spectra of three lines of CH$_3$CCH extracted from the PILS dataset convolved with a 24$''$ beam.\label{fig:convolve}} 
 \end{figure} 
\begin{table} 
\caption{Upper limit column densities ($N\rm{_{tot}}$) for the $^{13}$C and deuterated isotopologues of CH$_3$CCH, as well as the H$_2$ abundance ratio, in IRAS 16293B and IRAS 16293A. The 3$\sigma$ upper limits are determined using the excitation temperatures, FWHMs and H$_2$ column densities given in Table \ref{tab:coldens} for each source. }\label{tab:uplimits}
\centering
\footnotesize
\begin{tabular}{ccc}
\hline
\hline
Molecule&$N\rm{_{tot}}$  &\underline{$N\rm{_{tot}}$}\\
&(cm$^{-2}$)&$N$(H$_2$)\\
\hline
&IRAS 16293 A&\\
\hline
CH$_3^{13}$CCH & $<$2.44$\times$10$^{14}$ & $<$3.87$\times$10$^{-11}$ \\
$^{13}$CH$_3$CCH &$<$2.51$\times$10$^{14}$ & $<$3.98$\times$10$^{-11}$ \\
CH$_3$C$^{13}$CH &$<$2.51$\times$10$^{14}$ & $<$3.98$\times$10$^{-11}$ \\
CH$_3$CCD & $<$7.91$\times$10$^{14}$ & $<$1.26$\times$10$^{-10}$ \\
CH$_2$DCCH & $<$2.64$\times$10$^{14}$ & $<$4.19$\times$10$^{-11}$ \\
\hline
&IRAS 16293 B&\\
\hline
CH$_3^{13}$CCH &$<$2.44$\times$10$^{14}$ & $<$2.03$\times$10$^{-11}$ \\
$^{13}$CH$_3$CCH &$<$2.51$\times$10$^{14}$ & $<$2.09$\times$10$^{-11}$\\
CH$_3$C$^{13}$CH & $<$2.51$\times$10$^{14}$ & $<$2.09$\times$10$^{-11}$\\
CH$_3$CCD & $<$7.91$\times$10$^{14}$ & $<$6.59$\times$10$^{-11}$ \\
CH$_2$DCCH &$<$2.64$\times$10$^{14}$ & $<$2.20$\times$10$^{-11}$ \\
\hline

\end{tabular}
\end{table} 
\subsection{Spatial extent}\label{sec:spaex}

Figure \ref{fig:ch3cch_combmap} shows the spatial extent for four lines of CH$_3$CCH towards both sources. The top panels show the spectra of the lines and LTE best fit model towards IRAS 16293B and the bottom panel shows the same towards IRAS 16293A. Each line was chosen because it is a bright, unblended line. The middle panel is a velocity-corrected integrated emission (VINE) map of the line emission. A VINE map is a type of integrated emission map developed to determine the emission from molecules in regions where there are large velocity gradients and a large spectral line density. In such sources, it is difficult to determine a velocity interval to integrate over which does not include emission from other lines. VINE maps use a shifting velocity range to isolate the emission from only the line of interest over every pixel in the map. The shifting velocity range is determined using a velocity map of the source and the line width of the line. The velocity interval used is 3.2 km\,s$^{-1}$ and 1.2 km\,s$^{-1}$ for IRAS 16293 A and IRAS 16293B, respectively. Further details of how these maps are produced and a comparison between a VINE map and a standard integrated emission map for IRAS 16293 can be found in \citet{Calcutt2018b}. \\ 

\begin{table} 
\caption{Spatial extent of different transitions of CH$_3$CCH towards each hot corino in the IRAS 16293 system.}\label{tab:spaext}
\centering
\footnotesize
\begin{tabular}{ccc}
\hline
\hline
Transition&$E_{\rm up}$&Spatial extent$^{\dagger}$\\
\hline
&IRAS 16293A&\\
\hline
CH$_3$CCH 6 1 -- 5 1&\phantom{1}24\,K&1.7$''$ 233\,au\\
CH$_3$CCH 6 3 -- 5 3&\phantom{1}82\,K&1.5$''$ 214\,au\\
CH$_3$CCH 21 2 -- 20 2&218\,K&1.4$''$ 200\,au\\
CH$_3$CCH 20 3 -- 19 3&237\,K&1.3$''$ 185\,au\\
\hline
&IRAS 16293B&\\
\hline
CH$_3$CCH 6 1 -- 5 1&\phantom{1}24\,K&1.9$''$ 259\,au\\
CH$_3$CCH 6 3 -- 5 3&\phantom{1}82\,K&1.5$''$ 203\,au\\
CH$_3$CCH 21 2 -- 20 2&218\,K&1.1$''$ 148\,au\\
CH$_3$CCH 20 3 -- 19 3&237\,K&1.0$''$ 145\,au\\
\hline

\end{tabular}
\tablefoot{$^{\dagger}$The spatial extent in au is calculated for a source distance of 141\,pc as determined by \citet{Dzib2018}.}
\end{table} 

CH$_3$CCH traces the hot corino in both sources in the Band 3 and Band 7 datasets. Table \ref{tab:spaext} shows the spatial extent of CH$_3$CCH emission for each transition shown in Figure \ref{fig:ch3cch_combmap}. The values were calculated by determining the FWHM of a 2D-Gaussian fitted to the emission of each source. The emission that is observed is tracing scales smaller than the largest angular scale covered by the data. Lower upper energy level lines are tracing a slightly more extended scale than the higher upper energy level lines. The scale that this emission is tracing is comparable to the scale traced by other complex organic molecules in both A and B, including methyl cyanide (CH$_3$CN) and its isotopologues discussed in \citet{Calcutt2018b}, and methyl formate (CH$_3$OCHO, \citealt{Manigand2019}). \\

Recent analysis by \citet{Andron2018}, of CH$_3$CCH towards this source using IRAM 30\,m observations, claims CH$_3$CCH emission on scales of 1700 au. The Band 3 PILS dataset contains six of the same transitions as the IRAM data. The Common Astronomy Software Applications package (CASA; \citealt{Casa}) was used to convolve the PILS data with a 24$''$ beam, to determine how much of the emission from the hot corino components could explain the emission detected in the IRAM 30\,m observations. Figure \ref{fig:convolve} shows a spectrum of CH$_3$CCH lines, extracted from the convolved PILS data. The lines are heavily contaminated with absorption effects, however, comparing these lines to the spectra in Figure 4 of \citet{Andron2018}, the IRAM 30 m observations are $\sim$50\% brighter than the CH$_3$CCH emission found in the the PILS convolved observations. This suggests there is an additional cold component of CH$_3$CCH emission that is detected in the IRAM 30\,m observations, which is filtered out by the PILS dataset on scales larger than 2380 au. \\

\section{Chemical modelling} \label{sec:mod}

\begin{figure*} 
\begin{center} 
\includegraphics[width=18.5cm, angle=0, clip =true, trim = 0cm 0cm 1.7cm 0cm]{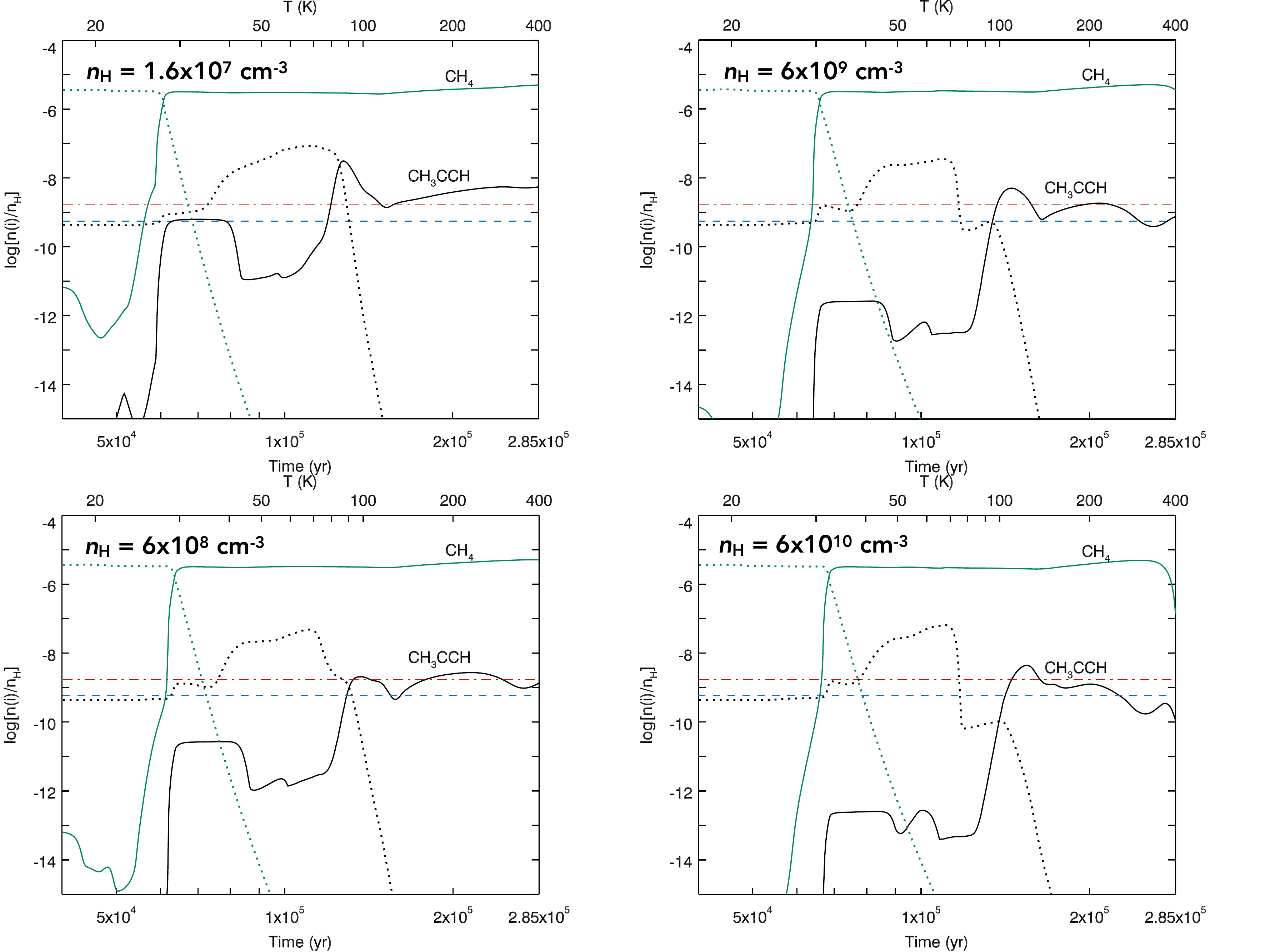}
\end{center} 
\caption{Abundances of CH$_3$CCH derived from chemical modelling for the warm-up stage of a hot core type model. Each panel represents a model with a different final collapse density, ranging from $n_{\mathrm{H}}= 1.6\times10^{7}$ -- 6$\times$10$^{10}$ cm$^{-3}$.  Solid lines denote gas-phase abundances, while dotted lines indicate grain-surface abundances. The red and blue dashed lines show the observational abundances in IRAS 16293A and B respectively.
 \label{fig:ch3cchmodel}} 
\end{figure*} 

To understand the results from this observational study in terms of the wider chemistry in IRAS 16293, chemical modelling of CH$_3$CCH was undertaken. The three-phase chemical kinetics model {\em MAGICKAL} \citep{Garrod2013, Willis2017} was used to model the formation and destruction pathways of CH$_3$CCH. It uses gas-phase, grain-surface and bulk ice reactions, using a chemical network based on that of \citet{Belloche2017}. A two-stage physical model is used. Initially, the cold collapse stage has an isothermal gas temperature of 10\,K, and the dust temperature cools from an initial value of 16\,K to a final value of 8\,K. This collapse takes 1.63$\times$10$^6$ years. This is then followed by a static warm-up to 400\,K. The chemical model is a single-point model, thus it has a uniform density. The initial abundances used in the code are given in Table \ref{tab:initialabun}. The cosmic-ray ionisation rate, $\zeta$, is assumed to be 1.3$\times$10$^{-17}$ s$^{-1}$, while the UV field has little effect during the hot-core stage due to high extinction. This model has been used previously to model CH$_3$CN and CH$_3$NC \citep{Calcutt2018a}, and NH$_2$CN in IRAS 16293 \citep{Coutens2018}. 

\begin{table} 
\caption{The initial fractional abundances with respect to the total hydrogen used in the three-phase chemical kinetics model {\em MAGICKAL.}\label{tab:initialabun}}\label{tab:initialabun}
\centering
\footnotesize
\begin{tabular}{cc}
\hline
\hline
Species&Abundance\\
\hline
He&9.00$\times10^{-2}$\\
N&7.50$\times10^{-5}$\\
O&3.20$\times10^{-4}$\\
H&2.00$\times10^{-3}$\\
H$_2$&4.99$\times10^{-1}$\\
C&1.40$\times10^{-4}$\\
S$^{+}$&8.00$\times10^{-8}$\\
Si$^{+}$&8.00$\times10^{-9}$\\
Na$^{+}$&2.00$\times10^{-8}$\\
Mg$^{+}$&7.00$\times10^{-9}$\\
P$^{+}$&3.00$\times10^{-9}$\\
Cl$^{+}$&4.00$\times10^{-9}$\\
\hline

\end{tabular}

\end{table} 

For this work, four models have been run to explore the solid-phase and gas-phase abundances of CH$_3$CCH at different final densities: $n_{\mathrm{H}}= 1.6\times10^{7}, 6\times10^{8}, 6\times10^{9}$,  and 6$\times$10$^{10}$ cm$^{-3}$, shown in Figure \ref{fig:ch3cchmodel}. These correspond to the varying densities seen in the IRAS 16293 system, with the highest density corresponding to the continuum peak of IRAS 16293B, and the lowest density corresponding to the density of the filament between IRAS 16293A and IRAS 16293B \citep{Jacobsen2018}. The subsequent warm-up phase starts at a dust temperature of 8 K, and reaches a final temperature of 400 K at $2.8\times10^{5}$ years. This timescale is used to represent an intermediate-timescale warm-up, where 2$\times$10$^{5}$ years is the time spent to reach a dust temperature of 200\,K. This is taken from \citet{Garrod2006}. 

CH$_3$CCH has both grain-surface and gas-phase formation routes in the modelling. On the grains, it is formed first through successive hydrogenation of smaller hydrocarbons. This process begins with hydrogenation of C$_3$. Hydrogenation of C$_3$ up to C$_3$H$_2$ proceeds without barrier, while hydrogenation of C$_3$H$_2$ has a small barrier of 250~K \citep{Garrod2013}. Further hydrogenation to CH$_3$CCH proceeds without a barrier. Reaction of CH$_3$CCH with H on the grain-surfaces has a yet-higher barrier of 1510~K, so very little CH$_3$CCH is further hydrogenated at low temperatures \citep{Tsang1992}. This is the primary destruction pathway for CH$_3$CCH on the grain-surface. At temperatures above 45\,K hydrogen abstraction from abundant radicals such as OH, NH$_2$, and CH$_2$OH also contributes to the destruction of CH$_3$CCH \citep{Dean1999}, leading to the formation of C$_3$H$_3$ and thus the formation of H$_2$O, NH$_3$ and CH$_3$OH.

In the gas-phase, the chemistry is more complex. CH$_3$CCH can form through neutral-neutral reactions, including C$_2$H$_4$ + CH $\rightarrow$ H + CH$_3$CCH \citep{Loison2017}. However, the primary gas-phase formation pathway used in the modelling is dissociative recombination of larger hydrocarbons (e.g. C$_3$H$_5^{+}$, C$_4$H$_5^+$). Some of these larger hydrocarbons are formed in part from CH$_3$CCH, so for some larger hydrocarbons this pathway becomes a feedback loop. However, for other larger hydrocarbons (as in the case of C$_3$H$_5^{+}$), they are formed almost exclusively from reactions of CH$_4$ with smaller hydrocarbons. The rates for these recombinations are taken from the OSU 2008 network \citep{Garrod2008}. Thus the increase in gas-phase abundance of CH$_3$CCH is directly tied to the desorption of CH$_4$ from the dust grains. CH$_4$ has a binding energy of 1300\,K \citep{Garrod2006}.

CH$_3$CCH is destroyed in the gas-phase by abundant ions (e.g. H$_3^{+}$, H$^{+}$, HCO$^{+}$). The rates for these ion-molecule are calculated according to the method of \citet{Herbst1986}. An additional contribution to the destruction of CH$_3$CCH in the gas-phase comes from atomic C at T > 40K. The atomic C produces a carbon insertion reaction, lengthening the carbon backbone and ejecting H/H$_2$ to form C$_4$H$_3$/C$_4$H$_2$, respectively. The rates for these reactions are taken from \citet{Harada2010}. 

The abundance of methane compared to the abundance of propyne in the different density models is shown in Figure \ref{fig:ch3cchmodel}. CH$_4$ desorbs from the grains at slightly higher temperatures in the high density models. In the lowest density model, it desorbs at about 26--27\,K, while in the highest density model, it desorbs at about 32 K.  This means that the low-temperature abundance peak is higher for CH$_3$CCH in the lower density models, as there is more CH$_4$ at earlier times in the gas-phase. 

In general, the grain-surface chemistry is what controls the final abundance of propyne. However, there are some key differences between the models. In the higher-density models, the grain-surface abundance of CH$_3$CCH decreases sharply before it desorbs from the grains. This is due to the larger abundance of atomic H on the grains at higher densities, which leads to more efficient hydrogenation of propyne to C$_3$H$_5$. After this point, propyne desorbs from the grains, and the change in gas-phase abundance of CH$_3$CCH is similar between the different models, all showing a slight dip in abundance at higher temperatures, which is more pronounced in the higher density model. This is due to propanal (C$_2$H$_5$CHO), which is slightly more abundant in the gas-phase in the lower-density models. Protonated propanal can recombine to form CH$_3$CCH. 

In the lower-density models, there is a larger amount of methane (CH$_4$), which is converted into larger hydrocarbons that then recombine and feed into the abundance of propyne. This is responsible for the difference in the level of abundance at $\sim$30 K in the models, and that heavily influences the later behaviour as well. This primarily has an impact on the gas-phase abundances in the models, whereas the peak grain-surface abundances do not vary significantly between the different models. Methane is predominantly formed on grains, through the hydrogenation of C. At higher temperatures, methane is also produced as a product of the CH$_3$ radical abstracting H atoms from species such as H$_2$ and H$_2$CO \citep{Baulch1992}. The CH$_3$ radical is formed primarily from dissociation of CH$_3$OH by cosmic-rays. The rate of dissociation is computed in the model to be 4.6$\times$10$^{-15}$ s$^{-1}$, assuming a canonical cosmic-ray ionisation rate of 1.3$\times$10$^{-17}$ s$^{-1}$ in our network.

\section{Discussion}\label{sec:diss}
The results from this study highlight how CH$_3$CCH emission in IRAS 16293A and IRAS 16293B is similar, both in spatial extent and abundance. To put this result in the wider context of chemistry in hot corinos/hot cores, it is important to compare it to similar studies in other sources. Whilst CH$_3$CCH has been detected in a variety of objects throughout the Galaxy, there are not many sources where the abundances have been determined with a high enough angular resolution, to ensure abundances are not heavily affected by beam dilution issues. One such high angular resolution study was performed recently of the high-mass hot core in the source AFGL 4176 by \citet{Bogelund2019}. Their study used ALMA observations to explore the continuum emission and the molecular content of AFGL 4176. The observed excitation temperatures and abundances of different molecules were then compared with a variety of sources, including IRAS 16293. In some cases, such as for CH$_3$CN, NH$_2$CHO, C$_2$H$_5$OH, CH$_3$OCH$_3$, CH$_3$OCHO and (CH$_2$OH)$_2$, abundances were found to be very similar between AFGL 4176 and IRAS 16293B.  However, for CH$_3$CHO, CH$_3$COCH$_3$, C$_2$H$_3$CN, C$_2$H$_5$CN, H$_2$CS, and SO$_2$ the abundances were notably different. The excitation temperatures found for these molecules in AFGL 4176 and IRAS 16293B also varied greatly, showing no correlation with the abundance similarities seen for some molecules. CH$_3$CCH was detected in the AFGL 4176 study but is 6 times more abundant relative to CH$_3$OH than in IRAS 16293B and 12 times more abundant relative to CH$_3$OH than in IRAS 16293A. It also has a higher excitation temperature of 320 K in AFGL 4176. Most notable, however, is the difference in spatial scale traced by CH$_3$CCH in AFGL 4176. The brightest continuum peak in the source, mm1, is coincident with the bulk of the complex organic molecule emission, with the exception of CH$_3$CCH. CH$_3$CCH has a double peaked morphology in the source, peaking both near the mm1 continuum peak position and at a second continuum peak, mm2, which is located north-west of mm1. A third continuum peak, mm3, is also seen in this source, located south-east of mm1. AFGL 4176-mm2 and -mm3 are located perpendicular to the major axis of mm1 and may be indicative of a large-scale outflow, which is consistent with the CO observations presented by \citet{Johnston2015}. This outflow could be responsible for the double-peaked morphology of CH$_3$CCH. Its emission, however, was not analysed towards mm3 to verify this possibility (B{\o}gelund, private communication).

\citet{Fayolle2015} performed an observational study of massive young stellar objects (MYSOs) with weak hot organic emission lines. Data from the IRAM 30 m and the Submillimeter Array (SMA), were used to determine molecular abundances towards three MYSOs, NGC 7538 IRS9, W3 IRS5, AFGL490, with known ice abundances, but without luminous molecular hot cores. They found that in contrast to other complex molecules, little or no CH$_3$CCH flux from the IRAM observations was recovered by the SMA observations, indicating an extended emission, which was spatially filtered by the SMA observations. This emission could have been large-scale emission and/or off-centred emission. 

The differences observed in the spatial emission of CH$_3$CCH compared to other molecules observed by both \citet{Fayolle2015} and \citet{Bogelund2019} suggest that CH$_3$CCH has a more complicated story than many other complex molecules. In contrast, its emission in IRAS 16293 does trace compact scales, tracing the hot corino components in both IRAS 16293A and IRAS 16293B up to scales of $\sim$2000 au, in addition to larger scale emission indicated by the IRAM 30\,m observations discussed in Section \ref{sec:spaex}. It is likely that the combination of both gas-phase and grain-surface pathways to the formation of CH$_3$CCH means that it is a `gateway' molecule tracing the interface between the hot corino region and the colder lower envelope, as well as tracing the larger colder envelope as is seen in the IRAM 30\,m observations.  

\section{Summary and conclusions}\label{sec:cons}
In this study, an analysis of CH$_3$CCH chemistry was undertaken, using high angular resolution ALMA observations towards the two forming protostars in the IRAS 16293 system. \\
Bright lines of CH$_3$CCH are detected towards both IRAS 16293A and IRAS 16293B, with upper energy levels ranging from 17 to 449 K. The excitation temperatures and abundances derived through LTE modelling, are similar in both sources. The 90--100\,K excitation temperatures are comparable to the excitation temperatures of several other molecules detected in these sources. 

These data show that CH$_3$CCH emission has a FWHM of $\sim$1--2$''$ ($\sim$145\,--\,260 au) around each source, indicating that this molecule is tracing the hot corino component in this dataset. This is similar to the spatial scale traced by other complex organic molecules, such as methyl cyanide and methyl formate in both sources. Lines with a lower upper energy level trace a slightly larger spatial scale.

The CH$_3$CCH emission found in this work was convolved with a 24$''$ beam, to determine whether the emission found in previous single dish studies in IRAS 16293, could be due solely to the two hot corino components. Spectra extracted from the convolved PILS dataset was 50\% weaker than previous IRAM 30\,m observations, indicating an additional cold component of propyne emission is present in the single-dish data. This cold component is not detected in the PILS dataset, which is likely due to it tracing larger scales ($>$2380 au) than the largest angular scale of the observations, and hence being filtered out. 

In addition to our observations, chemical modelling of CH$_3$CCH in IRAS 16293 was performed using the three-phase chemical kinetics model {\em MAGICKAL}. The modelling has found that to form enough CH$_3$CCH to match the observational abundances, a complex chemistry of both gas-phase and grain-surface reactions is required. On the grains, it is formed first through successive hydrogenation of smaller hydrocarbons. In the gas-phase, its primary formation pathway is dissociative recombination of larger hydrocarbons (e.g. C$_3$H$_5^{+}$, C$_4$H$_5^+$; \citealt{Garrod2013}). These larger hydrocarbons are formed through reactions of CH$_4$ with smaller hydrocarbons. Thus, the increase in gas-phase abundance of CH$_3$CCH is directly tied to the abundance of CH$_4$, which correlates strongly with the final collapse density of the model. 

The emission of CH$_3$CCH in IRAS 16293 was also compared to its emission in several high-mass star-forming regions. In the hot core AFGL 4176, the emission traces multiple continuum peaks in the data, which contrasts with the scale traced by other complex organic molecules detected towards the source.  In several other massive young stellar objects (MYSOs), with weak hot organic emission lines, CH$_3$CCH emission is found only on larger scales. The multiple scales traced by CH$_3$CCH are also seen in IRAS 16293, highlighting the complicated physical structures traced by this molecule. These differences are likely driven by a strong gas-phase and grain-surface chemistry leading to  abundant CH$_3$CCH formation in different environments.

CH$_3$CCH is an important complex organic molecule, whose presence in star-forming regions can be used to understand the temperature conditions around forming stars. This, combined with the large number of regions in both the Galaxy and beyond where it has been detected, offers an opportunity to compare how the chemistry varies across different objects, and explore the reasons for this variability. However, it is clear that the complicated chemistry that forms CH$_3$CCH allows it to be present on multiple scales. It is crucial that its physical structure is taken into account when determining its abundances and excitation temperatures, if it is to function as a benchmark molecule for different star-forming regions.

\begin{acknowledgements}
The authors would like to thank the referee for their helpful comments on the manuscript. The authors would like to acknowledge the European Union whose support has been essential to this research. In particular a European Research Council (ERC) grant, under the Horizon 2020 research and innovation programme (grant agreement No. 646908) through ERC Consolidator Grant ``S4F" to J.K.J. AC postdoctoral grant is funded by the ERC Starting Grant 3DICE (grant agreement 336474). S.F.W. acknowledges financial support from the Swiss National Science Foundation (SNSF) Eccellenza Professorial Fellowship PCEFP2\_181150. MND is supported by the Swiss National Science Foundation (SNSF) Ambizione grant 180079, the Center for Space and Habitability (CSH) Fellowship and the IAU Gruber Foundation Fellowship. This paper makes use of the following ALMA data: ADS/JAO.ALMA\#2013.1.00278.S and ADS/JAO.ALMA\#2012.1.00712.S. ALMA is a partnership of ESO (representing its member states), NSF (USA) and NINS (Japan), together with NRC (Canada) and NSC and ASIAA (Taiwan), in cooperation with the Republic of Chile. The Joint ALMA Observatory is operated by ESO, AUI/NRAO and NAOJ. This research has made use of NASA's Astrophysics Data System. This research has also made use of Astropy,\footnote{http://www.astropy.org} a community-developed core Python package for Astronomy \citep{astropy2013, astropy2018}.

\end{acknowledgements}

\bibliographystyle{aa}
\bibliography{ch3cch}

\end{document}